%% file: article.tex
\documentclass[10pt,a4paper]{elsarticle}

\def\eqbydef {~\stackrel{def}{=}~}

\usepackage{amsmath}    
\usepackage{graphicx}   
\usepackage{verbatim}   
\usepackage{color}      
\usepackage{subfigure}  
\usepackage{hyperref}   
\usepackage{float}
\usepackage{amssymb}

\DeclareTextSymbol{\degre}{OT1}{23}

\title{ {\bf Uncertainty in 2-point correlation function estimators and baryon acoustic oscillations detection in galaxy surveys} }  
     
\author[cea]{Antoine Labatie} 
\author[cea]{Jean-Luc Starck} 
\author[apc]{Marc Lachi\`eze-Rey}
\author[val1,val2]{Pablo Arnalte-Mur}

\address[cea]{Laboratoire AIM (UMR 7158), CEA/DSM-CNRS-Universit\'e Paris Diderot, IRFU, SEDI- SAP, Service d'Astrophysique, Centre de Saclay, F-91191 Gif-Sur-Yvette cedex, France.}
\address[apc]{Astroparticule et Cosmologie (APC), CNRS-UMR 7164, Universit\'e Paris 7 Denis Diderot 10, rue Alice Domon et L\'eonie Duquet F-75205 Paris Cedex 13, France.}
\address[val1]{Observatori Astron\`omic, Universitat de Val\`encia, Apartat de Correus 22085, E-46071 Val\`encia, Spain.}
\address[val2]{Departament	d'Astronomia	i	Astrof\'isica,	Universitat	de	Val\`encia,	46100-Burjassot, Val\`encia, Spain.}

\begin{document}

\input{epsf.tex}
\maketitle

\section*{Abstract}

\addcontentsline{toc}{section}{Abstract}

We study the uncertainty in different two-point correlation function (2PCF) estimators in currently available galaxy surveys. This is motivated by the active subject of using the baryon acoustic oscillations (BAOs) feature in the correlation function as a tool to constrain cosmological parameters, which requires a fine analysis of the statistical significance. 

We discuss how estimators are affected by both the uncertainty in the mean density  $\bar{n}$ and the integral constraint $\frac{1}{V^2}\int_{V^2} \hat{\xi}(r) d^3r =0$ which necessarily causes a bias. We quantify both effects for currently available galaxy samples using simulated mock catalogues of the Sloan Digital Sky Survey (SDSS) following a lognormal model, with a  Lambda-Cold Dark Matter ($\Lambda\text{CDM}$) correlation function and similar properties as the samples (number density, mean redshift for the $\Lambda\text{CDM}$ correlation function, survey geometry, mass-luminosity bias). Because we need extensive simulations to quantify small statistical effects, we cannot use realistic N-body simulations and some physical effects are neglected. 

Our simulations still enable a comparison of the different estimators by looking at their biases and variances. We also test the reliability of the BAO detection in the SDSS samples and study the compatibility of the data results with our $\Lambda\text{CDM}$ simulations.

\input{introduction}

\input{constraint}

\input{SDSSsamples}

\input{simulations}

\input{lognormal_2CF}

\input{conclusion}

\bibliographystyle{plain}
\bibliography{article}

\end{document}

%% file: introduction.tex
\section*{Introduction}
\addcontentsline{toc}{section}{Introduction}

The correlation function $\xi$ is the most popular tool for analyzing the distribution of galaxies \cite{peebles80}. 
Any model, like in particular the  standard $\Lambda\text{CDM}$, predicts a certain shape for $\xi(r)$ with a dependence on the cosmic parameters. Among the predictions, BAOs should imprint the matter correlation function. It is a relic of the sound waves in the early Universe when baryon and photons were coupled in a relativistic plasma before recombination which caused the wave propagation to end \cite{eisen98}. It can be seen as a small peak in the correlation function at a scale $r_s$ corresponding to the comoving distance of the sound horizon. 

The detection and localization of BAOs \cite{eisen05} give a confirmation of the cosmological paradigm and a tool to constrain cosmological parameters. The detection of BAOs in the Cosmic Microwave Background (CMB) provides the scale $r_s=153.3 \text{Mpc}$ and allows to constrain a combination of the Hubble constant $H(z)$ and comoving angular diameter distance $D_A(z)$ (see e.g. \cite{percival10},\cite{kazin10}). Further, using the value of $\Omega_m h^2$, also well constrained by CMB measurements, the BAO scale restricts the preferred regions for $\Omega_m$ and $h$. 

The main difficulty for detecting and analyzing BAOs in large scale structures comes from the low statistical significance of the signal. It can only be seen on the widest redshift surveys, and has been most significantly detected in samples including Luminous Red Galaxies (LRGs). In addition to the statistical uncertainty the signal is affected by observational effects that may not be taken into account correctly, such as redshift distortions, scale-dependent mass-luminosity bias in the population of galaxies or wrong redshift to distance conversion. 

We will not study these systematic effects; instead we focus on the statistical uncertainty in the BAO signal estimation through correlation functions. There are two types of statistical uncertainties. The first one comes from cosmic fluctuations  due to limited sample volume, and the other one from the finite number of galaxies which do not trace exactly the underlying field (i.e. shot noise). 

There are various estimators of the correlation function. Their bias expresses the difference between their expected value and the value of the physical quantity of concern. Estimators are also subject to variance. In practice there is no way to evaluate the bias of the estimator if it exists, and it must be considered itself as a source of uncertainty, in addition to the estimator's variance. 

Usual criterions to compare statistical estimators involve both the variance and the bias. For example, when measuring the quality in terms of mean-squared error, biased estimators could outperform unbiased ones. This is the well-known bias-variance tradeoff that depends on the way we measure the quality of estimation. For some cosmological analysis, the presence of a bias could be problematic if not taken into account. For example, fitting model correlation functions to the data, taking only into account the covariance matrix and not the bias, would lead to a false estimation of confidence intervals for the model parameters. 

For our study, we use simulations with $\Lambda\text{CDM}$ power spectrum on the same volume as the data and with the same estimated parameters (density of galaxies, mass-luminosity bias, mean redshift for the power spectrum). Our simulations assume a lognormal model (described in section \ref{simulations}) for the density field as proposed by \cite{cj91}, which has proven to be valid for a good range of scales. The model used has physically motivated features, although it is not entirely realistic. It does not completely take into account the systematic effects mentioned above: redshift distortions, scale-dependent mass-luminosity bias in the population of galaxies, wrong redshift to distance conversion.

There have been several studies to compare the different estimators of the correlation function (\cite{pons99}, \cite{ker00}). Here we perform similar comparisons for current galaxy surveys, focusing on large-scale effects and BAO detection. We arrive at similar conclusions as previous studies when ranking estimators in terms of performance. Our second goal is more specific, focusing on the bias caused by the integral constraint for correlation function estimation. Such a bias is expected for all sizes of survey in a fractal Universe \cite{lab98} and below the scale of homogeneity in the standard cosmological model. We study whether this systematic alters the estimation or can be neglected in current galaxy surveys, in particular for BAO study.

The plan of the paper is as follows. In \ref{2PCF_estimators} we present the different estimators of the correlation function that we consider. We recall some of their properties, in particular their sensitivity to the uncertainty in the mean density $\bar{n}$ in \ref{sensitivity}, and the bias imposed by the integral constraint in \ref{constraint}. In \ref{samples} we present the SDSS samples that we want to mimic with our simulations (one LRG sample and one main sample), and in \ref{simulations} the lognormal model and our procedure for fitting simulation parameters to the data. Finally in \ref{lognormal_uncertainty} we perform the analysis of the uncertainty in the $\xi$ estimation. We compare the quality of the different estimators in \ref{comparison}, and look at the effect of the integral constraint in the simulations in \ref{constraint_data}. In \ref{reliability} we look at the reliability of the BAO detection in the SDSS LRG and main samples, and see in \ref{compatibility} if the $\xi$ estimated on the data is compatible with our lognormal simulations with a given $\Lambda\text{CDM}$ power spectrum.

%% file: constraint.tex
\section{2PCF estimators and bias}
\label{bias}

\subsection{2PCF estimators}
\label{2PCF_estimators}
The two-point correlation function is a second order statistic that describes the clustering of a field or a point process. More precisely $\xi(r)$ measures the excess of probability to find a pair of points in two volumes $dV_1$ and $dV_2$ at distance $r$ compared to a random distribution.

\begin{equation}
dP_{12}=\bar{n}^2 [1+\xi(r)] dV_1 dV_2
\end{equation}

where $\bar{n}$ is the expected density of the distribution.   \\

Computing the correlation function requires to have a 3D map of galaxies. In practice galaxies are located using their angular position on the sky and their distance from the observer. The distance of the galaxy is obtained indirectly by its redshift, which can be measured with high precision using spectroscopy. Assuming a cosmological model, the distance of the galaxy is obtained using the relation from redshift to distance (for this cosmological model).

There are various estimators of the correlation function, most using random catalogues with identical geometry to measure this excess of probability. Let $N_D$ and $N_R$ be the number of galaxies respectively in the data and random catalogues. We define $DD(r)$, $RR(r)$ and $DR(r)$ as the number of pairs at a distance in $[r \pm dr/2]$ of respectively data-data, random-random and data-random galaxies. We also define $N_{DD}$, $N_{RR}$ and $N_{DR}$ as the total number of corresponding pairs in the (real or random) catalog. With the convention of counting pairs only once we have:

\begin{eqnarray}
N_{DD} & = & \frac{N_D (N_D-1)}{2} \\
N_{RR} & = & \frac{N_R (N_R-1)}{2} \\
N_{DR} & = & N_R N_D \\
\end{eqnarray}

In this paper we will use 4 different estimators, Peebles-Hauser \cite{ph74}, Davis-Peebles \cite{davis83}, Hamilton \cite{hamilton93} and Landy-Szalay \cite{landy93}, which have the following expressions:

\begin{eqnarray*}
\hat{ \xi}_{PH}(r) & = &  {N_{RR} \over N_{DD}} {DD(r) \over RR(r)} -1 \\
\hat{\xi}_{DP}(r) & = &  {N_{DR} \over N_{DD}}  {DD(r) \over DR(r)} -1 \\
\hat{\xi}_{HAM}(r) & = &   {{N_{DR}}^2 \over N_{DD} N_{RR}}  {DD(r) RR(r) \over [DR(r)]^2} -1 \\
\hat{\xi}_{LS}(r) & = &  1 + {N_{RR} \over N_{DD}} {  DD(r) \over  RR(r)}  -  2 {N_{RR} \over N_{DR}}  { DR(r) \over RR(r)}
\end{eqnarray*}

Estimating $\xi$ would be easier knowing the exact number of points in the volume expected from the distribution. In practice we can only estimate it with the empirical quantities $N_D$ and $N_{DD}$. We show in section \ref{sensitivity} that Hamilton and Landy-Szalay only depend on the second order on this uncertainty in the mean density, and thus perform better. Moreover in \cite{landy93} Landy-Szalay has been proven to be nearly of minimal variance for a random distribution (i.e. Poisson with no correlation).

\subsection{Uncertainty in the mean density}
\label{sensitivity}
We show the calculations given in \cite{hamilton93} in a simple case where the sample is volume-limited (i.e. with a constant expected density in the sample), so that the optimal strategy is to weight all galaxies equally. The empirical density in the catalogue $n$ is a sum of Dirac functions on the galaxies of the catalogue. If $\bar{n}$ is the expected density then $\delta$ is the relative fluctuation in the sample:

\begin{equation}
\delta = \frac{n-\bar{n}}{\bar{n}}
\end{equation}

We write $W$ the indicator function of the sample volume and $\langle . \rangle$ the integration on the volume. For example $\langle  W({\bf x}) \, n({\bf x}) \rangle$ is the integration of the empirical density and thus equals the number of points in the sample. We introduce the following quantities (with $\bar{\delta}$ and $\Psi$ that have statistical expectations of 0): 

\begin{equation}
\bar{\delta} = \frac{\langle  W({\bf x}) \, \delta({\bf x}) \rangle}{\langle W({\bf x}) \rangle}
\end{equation}

\begin{equation}
\Psi(r) = \frac{\langle \delta({\bf x}) \, W({\bf x}) W({\bf y}) \rangle_r} {\langle W({\bf x}) W({\bf y}) \rangle_r}
\end{equation}

\begin{equation}
\hat{\xi}(r) =\frac{\langle \delta({\bf x}) \delta({\bf y})  \, W({\bf x}) W({\bf y}) \rangle_r} {\langle W({\bf x}) W({\bf y}) \rangle_r}
\label{xihat}
\end{equation}

where $\langle . \rangle_r$ means a double integration in the volume, restricted to ${\bf x}$ and ${\bf y}$ separated by a distance in $[r \pm dr/2]$. $\hat{\xi}$ is an unbiased estimator of the real $\xi$ but we cannot calculate it since we do not know $\bar{n}$ and $\delta$. 

With short calculations it is possible to express the different estimators with the quantities  $\hat{\xi}$, $\bar{\delta}$ and $\Psi$ \cite{hamilton93}:

\begin{equation}
\hat{\xi}_{PH}(r) = \frac{\hat{\xi}(r) + 2\Psi(r) - 2\bar{\delta} - \bar{\delta}^2} {[1+\bar{\delta}]^2 }
\label{Srewrite}
\end{equation}

\begin{equation}
\hat{\xi}_{DP}(r)  = \frac{\hat{\xi}(r) + \Psi(r) - \bar{\delta} - \Psi(r) \, \bar{\delta}} {[1+\bar{\delta}] \left[ 1+\Psi(r) \right]}
\label{DPrewrite}
\end{equation}

\begin{equation}
\hat{\xi}_{H}(r) = \frac{\hat{\xi}(r) - \Psi(r)^2} {[1+\Psi(r)]^2}
\label{Hrewrite}
\end{equation}

\begin{equation}
\hat{\xi}_{LS}(r) = \frac{\hat{\xi}(r) - 2\bar{\delta} \Psi(r) + \bar{\delta}^2} {[1+\bar{\delta}]^2 } \\
\label{LSrewrite}
\end{equation}

These formulas explain the superiority of Hamilton and Landy-Szalay estimators, with $\Psi$ and $\bar{\delta}$  terms at the second order in the numerator. Terms in the denominator are not important since they generate a small relative error, whereas terms in the numerator can generate a high relative error when their values become non negligible compared to $\hat{\xi}$. For Hamilton and Landy-Szalay estimators, the error is dominated by the one of $\hat{\xi}$ and not really affected by $\Psi$ and $\bar{\delta}$, which are linked to the uncertainty in $\bar{n}$. 

With these formula we see that the estimators are biased in the general case. Indeed $\bar{\delta}$ and $\Psi(r)$ have expected value 0 and $\hat{\xi}(r)$ has expected value $\xi(r)$, but the terms are combined in multiplications and divisions. So we do not get the expected value of the left-hand side by replacing each term by its expected value in the right-hand side of equations (\ref{Srewrite}),  (\ref{DPrewrite}), (\ref{Hrewrite}), (\ref{LSrewrite}).


\subsection{The integral constraint}
 \label{constraint}
 
The random catalogue is used to measure an excess of pairs compared to a random distribution. Equivalently it can be seen as a tool to calculate volumes. Let $V$ be the domain of the sample, if we take the limit $N_R \rightarrow \infty$:

\begin{eqnarray}
    \frac{RR(r)}  {N_{RR}} 					              		    & = &  \frac{ \text{\# pairs at distance } r'\in [r \pm dr/2]}  { \text{ \# pairs} } \nonumber \\
f(r) 
\eqbydef
  \lim \limits_{N_R \rightarrow \infty}   \frac{RR(r)}  {N_{RR}}    & = &   \frac{1} {|V|^2} \int_V d^3 {\bf x} \int_V d^3 {\bf y} \hspace{2mm} 1_{ |{\bf y-x}| \in [r \pm dr/2] } 
      \label{f}
\end{eqnarray}

To simplify the text we define $I$ and $\hat{I}_{PH}$, $\hat{I}_{DP}$, $\hat{I}_H$, $\hat{I}_{LS}$ ($\hat{I}$ when refering to any estimator) as the values of the integration against $f(r)$ for the real correlation and for the different estimators:

\begin{eqnarray}
I & \eqbydef & \int_0^{r_{\text{max}}} f(r) \xi(r) \\
\hat{I}_{PH} & \eqbydef & \int_0^{r_{\text{max}}} f(r) \hat{\xi}_{PH}(r)
\end{eqnarray}

\noindent with $r_{\text{max}}$ the maximum distance between 2 points in the volume. \\

We will show that there is a constraint on the Peebles-Hauser estimator $\hat{\xi}_{PH}(r)$ imposing the following equality, regardless of the real function $\xi(r)$ that is estimated:

\begin{equation}
\hat{I}_{PH} =0
\label{sampleconstraint}
\end{equation}

For a smooth sample and small separation $r$, the inner integral in equation (\ref{f}) equals for nearly all $\bf{x}$ the volume of the spherical envelope 
${\bf y} \in V_r $ with $|{\bf y}-{\bf x}| \in [r \pm dr/2]$. So for small $r$ we get $f(r) \approx \frac{|V_r|}{|V|} = \frac{4 \pi r^2 dr}{|V|}$, and if it was
the case for all $r$ the constraint (\ref{sampleconstraint}) would become:

\begin{equation}
 \int_{\mathbb{R}^3} \hat{\xi}({\bf r}) d^3 {\bf r}=0.
 \label{usualconstraint}
 \end{equation}
  
But when $r$ becomes non negligible compared to the sample size, $f(r) \neq \frac{|V_r|}{|V|}$, and so the constraint (\ref{sampleconstraint}) is different from (\ref{usualconstraint}) and depends on the sample volume and geometry.

Let us show the relation (\ref{sampleconstraint}) for the Peebles-Hauser estimator:

\begin{equation}
\hat{\xi}_{PH}(r) =  {N_{RR} \over N_{DD}} {DD(r) \over RR(r)} -1 \approx { 1 \over f(r)}  { 1 \over N_{DD}}   DD(r) -1 \\
\end{equation}

In practice the integration consists in making the sum over all bins $r_i$ of the correlation function estimated up to $r_{max}$:

\begin{eqnarray*}
\hat{I}_{PH}  = \sum_i f(r_i) \hat{\xi}_{PH}(r_i) & \approx &  \sum_i  f(r_i)  \left[ { 1 \over f(r_i)} {1 \over N_{DD}}  DD(r_i) -1 \right]  \\
& = &   {1 \over N_{DD}} \sum_i DD(r_i) - \sum_i f(r_i) = 1-1 =0 \\
\end{eqnarray*}

It is possible to show that the same constraint  $\hat{I}=0$  is approximately verified for the other estimators. For this we need to simplify $DR(r)$  in the limit $N_R \rightarrow \infty$. 
 
 \begin{eqnarray}
 \frac{1}{N_D N_R}  DR(r) =   \frac{1}{N_D} \sum_{{\bf d_j}}  \frac{ \text{\# random points ${\bf r_i} $ s.t. }  |{\bf r_i-d_j}| \in [r \pm dr/2]}  { \text{\# random points } }\nonumber \\
g(r)=  \lim \limits_{N_R \rightarrow \infty} \frac{1}{N_D N_R}  DR(r) =   \frac{1}{N_D} \sum_{{\bf d_j}}  \frac{1}{V} \int_V d^3 {\bf y} \hspace{2mm} 1_{ |{\bf y-d_j}| \in [r \pm dr/2] }  
\label{g} 
\\ \nonumber
\end{eqnarray}
 
This functions $g(r)$ depends on the point positions in the catalogue. We can make another approximation if the size of the correlation is small compared to the volume and if there are enough data points. Then data points are approximately uniformly distributed in the volume, and we can replace the mean on data positions by the mean on the volume:

 \begin{equation}
g(r)  \approx  \frac{1}{V} \int_V d^3 {\bf x}  \left( \frac{1}{V}  \int_V d^3 {\bf y} \hspace{2mm} 1_{ |{\bf y-x}| \in [r \pm dr/2] } \right) =f(r)
\end{equation}

Under this approximation all estimators are equivalent and verify the integral constraint. But the last approximation is not as good as for Peebles-Hauser estimator, and the constraint should be less tight. 

We see again that the estimators are biased in the general case. The real correlation function need not satisfy the constraint, whereas estimators do (approximately) verify it and thus their expected value also.

\subsection{Effect of the integral constraint on the bias}

To show how the constraint can affect the correlation function estimation we generated realizations of segment Cox process (see \cite{martsaar02}). The field consists in segments of length $l$ randomly distributed in the volume and points randomly distributed on each segment. The intensity of the point process $\lambda$ is equal to the mean length of segments per unit volume $L_V$ times the mean number of points per unit length $\lambda_l$. This process is easy to sample and its correlation function is known analytically \cite{stoyan95}:

\begin{equation}
\xi(r) = \left\{
    \begin{array}{ccc}
        \frac{1}{2 \pi r^2 L_V} - \frac{1}{2 \pi r l L_V} & \text{for} &  r\leq l \\
        0 & \text{for} &  r \geq l
    \end{array}
\right.
\end{equation}

\begin{figure}[H]
 \begin{center}$
 \begin{array}{cc}
\includegraphics[scale=0.28]{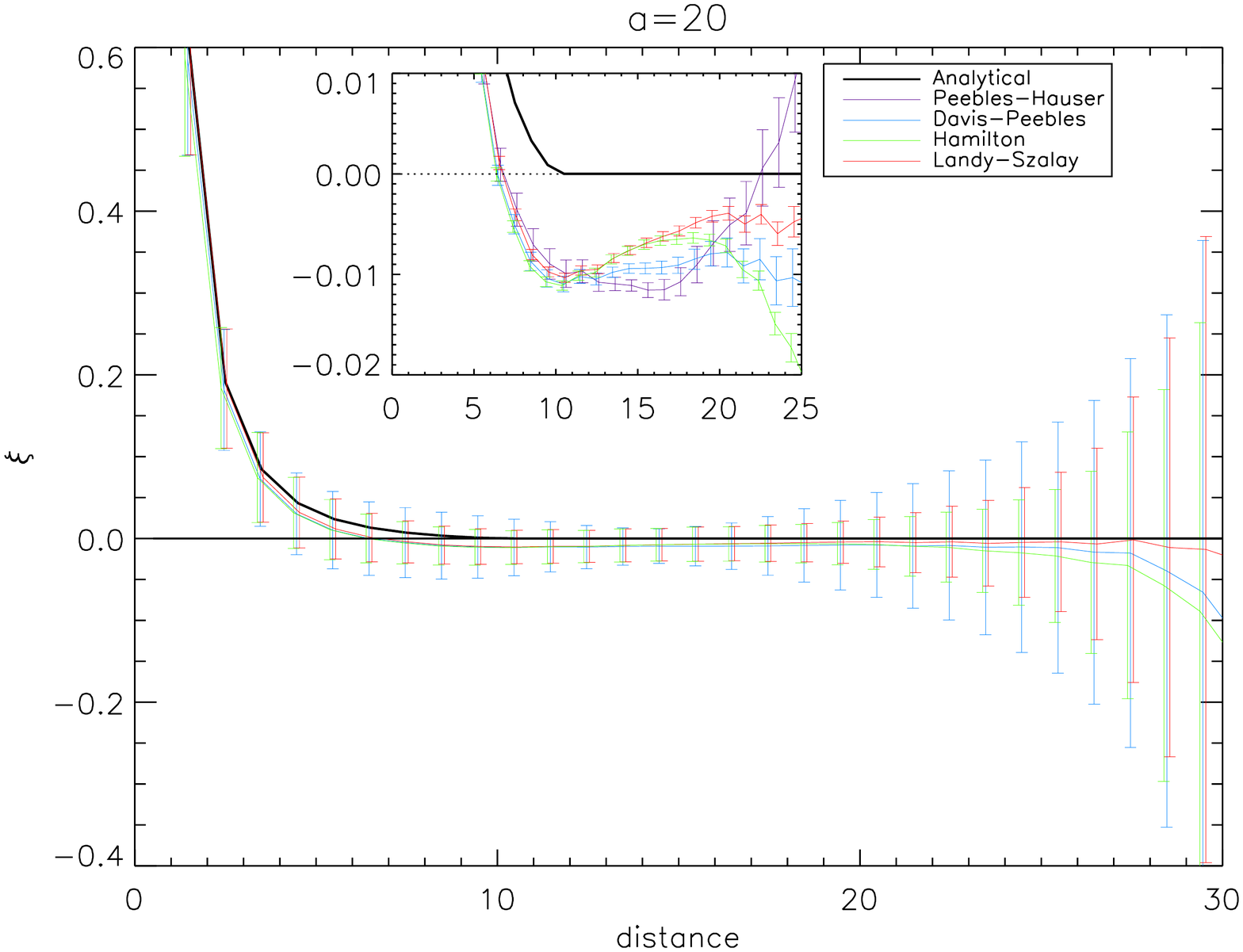}
\includegraphics[scale=0.28]{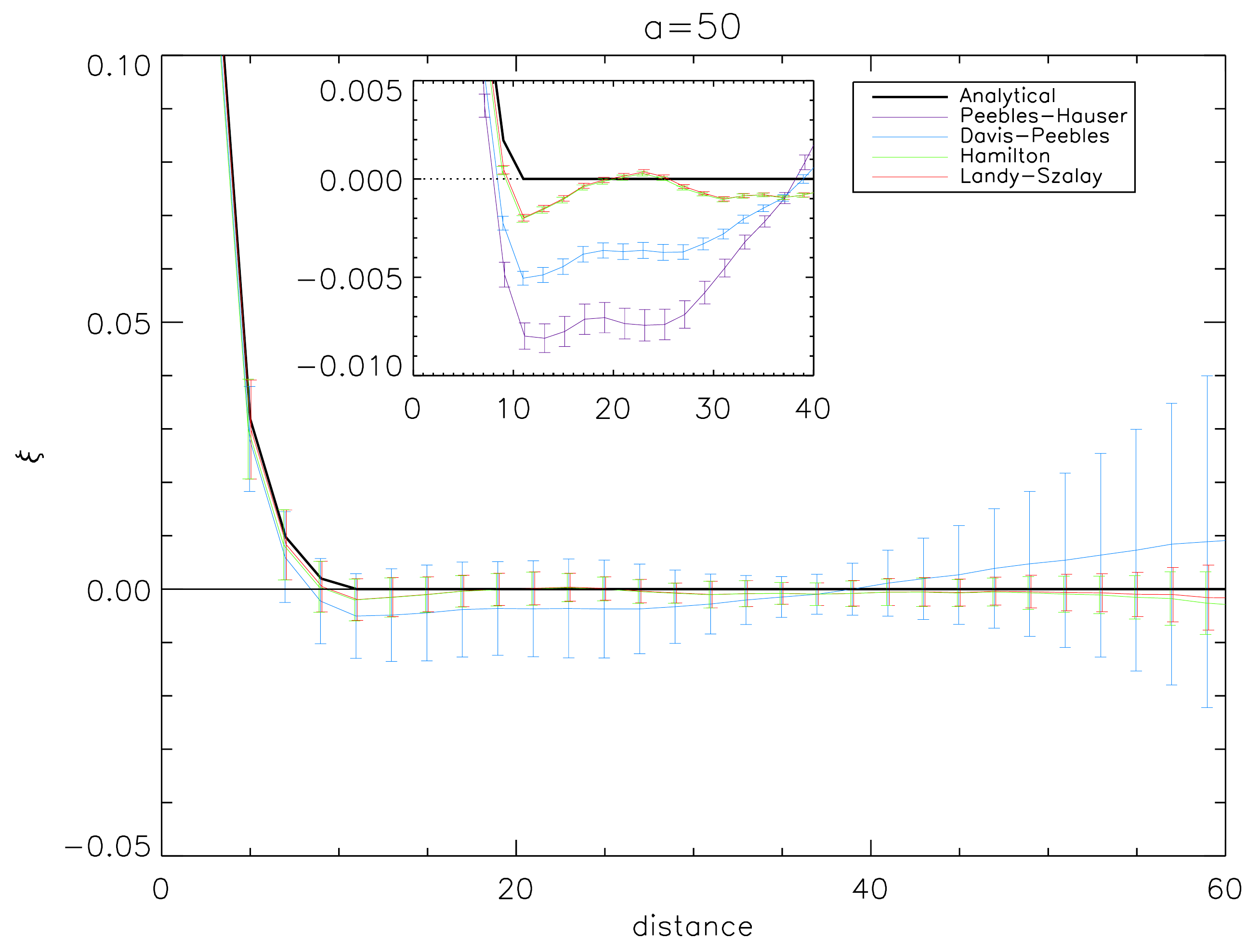}
\end{array}$
\end{center}
\caption{Mean and $\hat{\sigma}$ for the different estimators on $N=2000$ Cox realizations for cube size $a=20$ (left panel) and $N=512$ realizations for $a=50$ (right panel). We plot the analytic function (black), Peebles-Hauser (purple), Davis-Peebles (light blue), Hamilton (green), Landy-Szalay (red). In inset we zoom over the biased region with error bars $\frac{\hat{\sigma}}{\sqrt{N}}$ which is approximately the standard deviation of the mean on $N$ realizations.}
\label{CoxEstimators} 
\end{figure}

It is always non-negative so the integral constraint forces false negative values for the estimators. We considered the process with segment length $l=10$ (units here are arbitrary), a mean length by unit volume $L_V=0.1$ and a mean number of points per unit length $\lambda_l=1.8$. We calculated the correlation function estimators on $2000$ cubes of sizes $a=10$, $2000$ cubes of size $a=20$ and $512$ cubes of size $a=50$ . We plot figure \ref{CoxEstimators} the mean value of the estimators on the samples and the empirical $\hat{\sigma}$ value. To exemplify the presence of the bias we show in the insets the empirical $\hat{\sigma}$ divided by $\sqrt{N}$, with $N$ the number of realizations, which gives the uncertainty in the empirical mean. A difference between the empirical mean and the real $\xi$ much larger than $\frac{\hat{\sigma}}{\sqrt{N}}$ means a bias is present in the estimators. 

We observe that a bias is present for all estimators and for all sizes of cubes. As expected it becomes smaller when the sample size increases just like the variance decreases. For Landy-Szalay and Hamilton estimators the bias also decreases faster than the estimators' variances. The bias approximately equals half of the standard deviation $\hat{\sigma}$ in a large region for $a=10$ and $a=20$, whereas it is very small compared it for $a=50$. Biases are similar for the different estimators for $a=10$ and $a=20$, although Landy-Szalay and Hamilton have smaller variances than Peebles-Hauser and Davis-Peebles.

\begin{figure}[H]
 \begin{center}$
 \begin{array}{cc}
\includegraphics[scale=0.3]{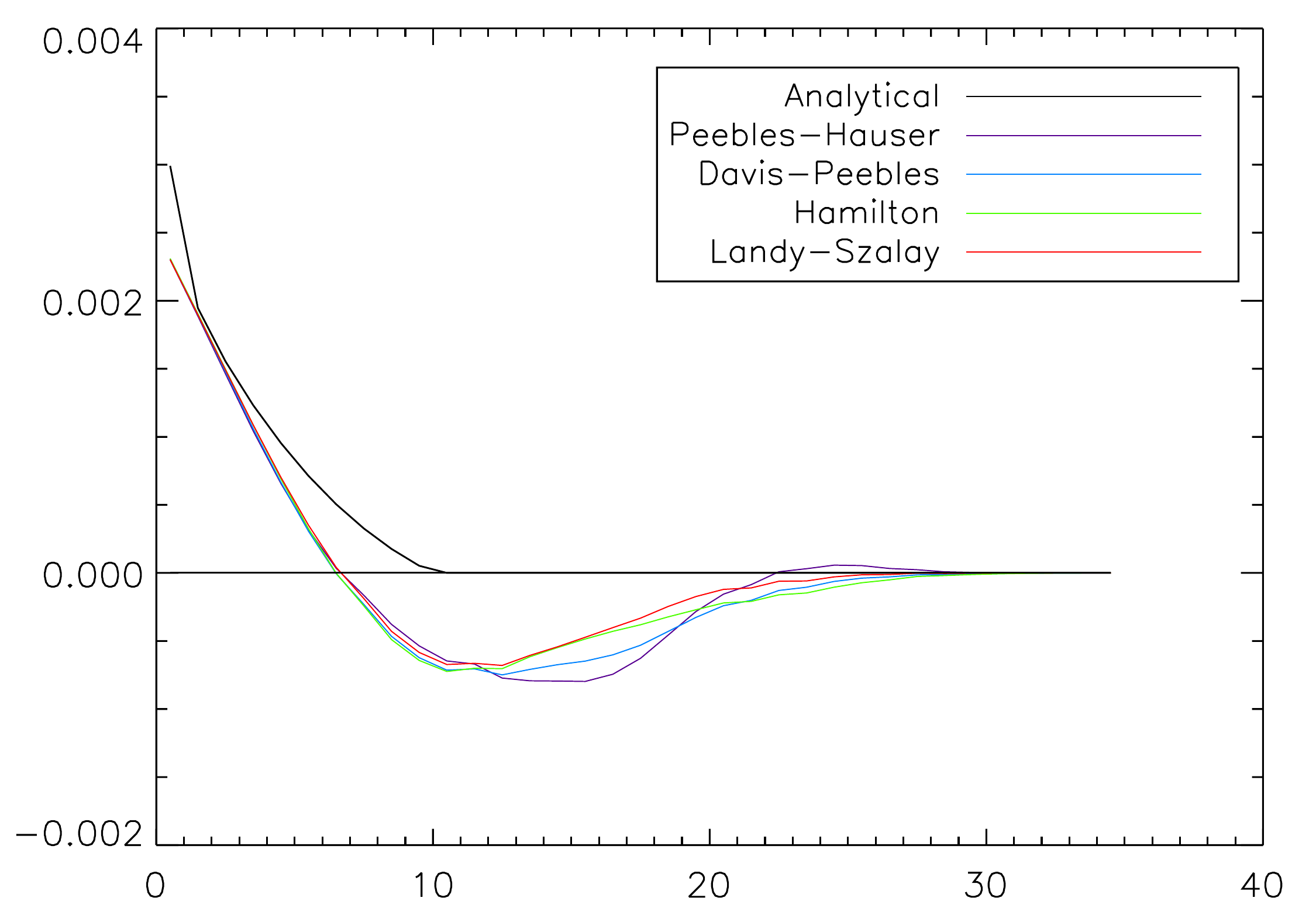}
\includegraphics[scale=0.3]{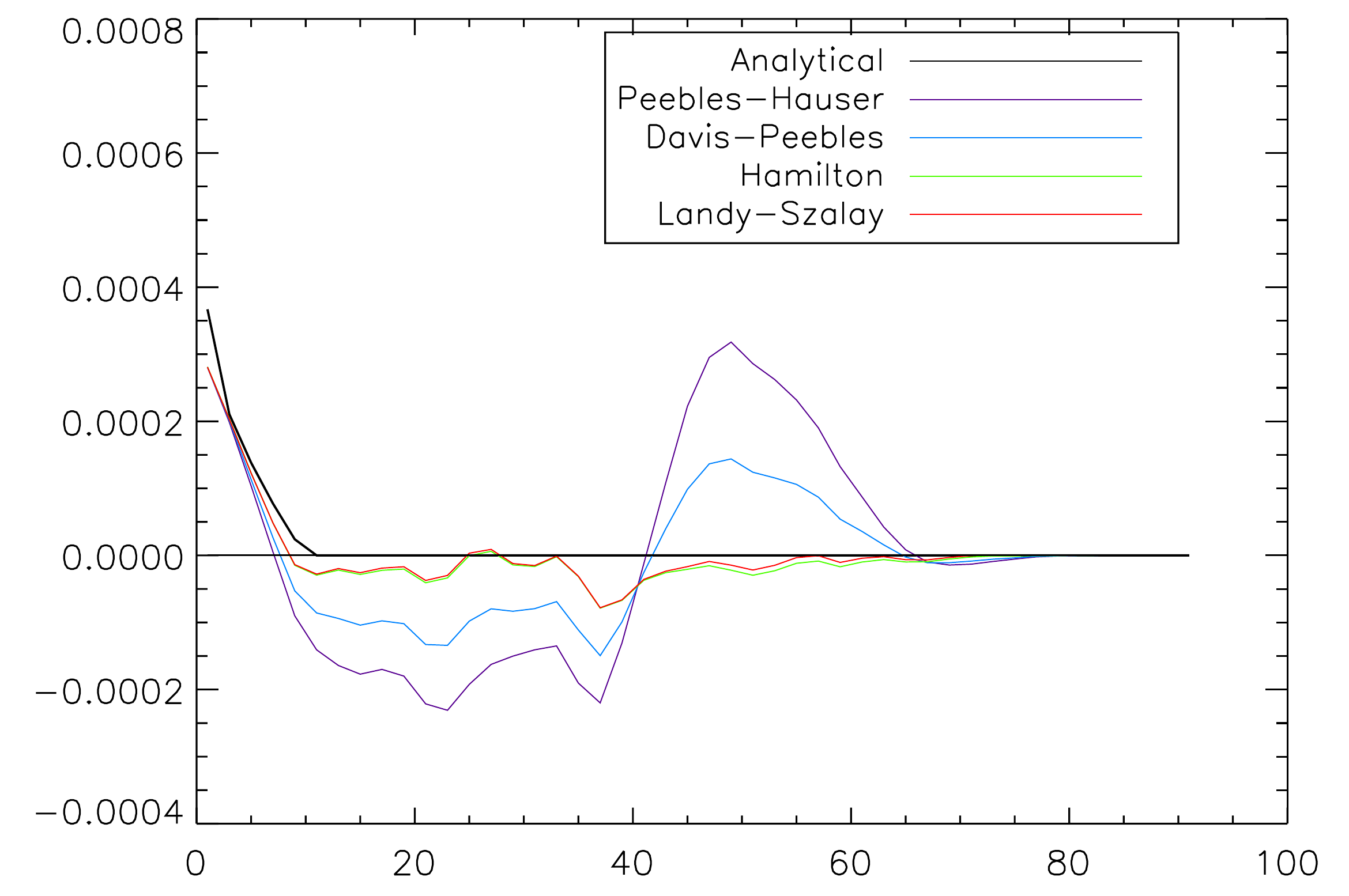}
\end{array}$
\end{center}
\caption{Weighted estimators $f(r_i) \hat{\xi}(r_i)$ for $a=20$ (left panel) and $a=50$ (right panel). We plot it for the analytic function (black), Peebles-Hauser (purple), Davis-Peebles (light blue), Hamilton (green), Landy-Szalay (red).}
\label{CoxwEstimators} 
\end{figure}

The effect of the bias is to force negative values at intermediate scales, so that the weighted sum in $\hat{I}$ approaches 0. Figure \ref{CoxwEstimators} shows the weighted estimators $f(r_i) \hat{\xi}(r_i)$ and how the integral cancels for the estimators and not for the real $\xi$. The effect is clear for $a=20$ and $a=10$ (not shown because results for $a=10$ and for $a=20$ have similar trends). However for $a=50$, the bias comes not entirely from the integral constraint as the weighted function takes alternatively negative and positive values. So the small bias that is still present could come from other effects (e.g. finite number of random points).

We show table \ref{constraintCox} the value of $I$ for the real $\xi$ and $\hat{I}$ for the estimators' means. The constraint is nearly satisfied ($\hat{I} \approx 0$), especially for Peebles-Hauser, even when the real $\xi$ does not verify it ($I \gg \hat{I}$).

The weight function $f$ sums to 1 (see equation (\ref{f})), so a the difference in $\sum_i f(r_i) \xi(r_i)$ and $\sum_i f(r_i) \hat{\xi}(r_i)$ ($I$ and $\hat{I}$) implies  in average a similar  difference between $\xi$ and $\hat{\xi}$. Negative bias may compensate positive bias in the integral, so it can be an underestimation. 

For the Landy-Szalay and Hamilton estimators the constraint gets weaker between $a=20$ and $a=50$. These values of $a$ correspond to values of $I$ for the real $\xi$ of approximately $0.01$ and $0.001$. A quantity which is more intuitive than $I$ is the normalized mass variance inside a sample $V$:

\begin{equation}
\sigma^2(V)= \frac{ \mathbb{E}\left[M(V)^2\right] -\mathbb{E}\left[M(V)\right]^2} { \mathbb{E}\left[M(V)\right]^2 }
\end{equation}

$\sigma(V)$ represents the fluctuation of mass in the sample. It can be shown that $I$ is equal to $\sigma^2(V)$ up to the shot noise variance (see \cite{gabrielli02}), which can be usually neglected. Thus we can express conditions for the constraint to be weak or negligible in terms of the $\sigma(V)$ value. The cubic samples with $a=20$ and $a=50$ correspond respectively to $\sigma(V) \approx 0.1$ and $\sigma(V)\approx 0.03$. So the constraint still affects the estimation for a $10\%$ homogeneity level and starts to be weak for a $3\%$ homogeneity level.

\begin{table}[H]
\begin{center}
\begin{tabular}{|c|c|c|c|c|c|c|}
\hline    & $I$ & $\sigma(V)$ & $\hat{I}_{PH}$ & $\hat{I}_{DP}$ & $\hat{I}_{H}$ & $\hat{I}_{LS}$\\ 
\hline
$a=10$ &  $0.059$ & $0.24$  & $2.87 \times 10^{-5}$ &  $9.24 \times 10^{-5}$ & $7.16 \times 10^{-3}$ & $0.0125$ \\
\hline
$a=20$ & $9.6 \times 10^{-3}$  & $0.098$  &  $1.88 \times 10^{-5}$ & $ -5.54 \times 10^{-4}$ & $2.23 \times 10^{-4}$ & $1.47 \times 10^{-3}$ \\
\hline
$a = 50$ & $7.8 \times 10^{-4}$ & $0.027$  &  $3.8 \times 10^{-6}$ &  $-7.04 \times 10^{-5}$ & $-1.86 \times 10^{-5}$ & $1.02 \times 10^{-4}$ \\
\hline
\end{tabular}
\caption{Values of $I$ and $\hat{I}$ for different estimators on cube sizes $a=10$, $a=20$ and $a=50$}
\label{constraintCox}
\end{center}
\end{table}

%% file: SDSSsamples.tex
\section{Samples and simulations}
\subsection{SDSS galaxy samples}
\label{samples}

We want to test the reliability of the correlation function estimation on current galaxy surveys. 
The largest survey up to date is the SDSS with a final version in Data Release 7 (DR7,\cite{aba08a}). 
It contains a magnitude-limited sample of galaxies (main) and a nearly-volume-limited sample of LRGs. 
For all catalogues we adopt a $\Lambda\text{CDM}$ cosmological model with $\Omega_M=0.27$, and $\Omega_\Lambda=0.73$.

To create volume-limited samples of the main we use the catalogue available in Mangle's webpage\footnote{http://space.mit.edu/$\sim$molly/mangle/}. This catalogue is based on the New York University Value-Added Galaxy Catalog \cite{nyuvagc}. It contains $r$-band absolute magnitudes ($M_r$) for each galaxy that are already $K$-corrected 
and corrected for evolution at a fiducial redshift of $z=0.1$ following \cite{bla03b}. The $K$-correction and evolution correction are required because galaxies are observed at different redshifts. The $K$-correction converts galaxy spectra from observed to emitted frame \cite{hogg02}. Evolution correction is required to take into account the time-evolution of galaxies (and thus their spectra) from their individual observed redshift to a common redshift for all galaxies \cite{fioc97}.
Comoving distances and absolute magnitudes are given in the Mangle cosmology, so we convert them in the cosmology we use ($\Omega_M=0.27$, $\Omega_\Lambda=0.73$). 

We also use a volume-limited sample of LRGs drawn directly from SDSS-DR7. LRGs are early-type galaxies
selected using different luminosity and colour cuts \cite{eisen01a}, and extending to higher redshift. We compute
the $K$-corrected $g$-band absolute magnitudes ($M_g$), and corrected for evolution at a fiducial redshift of $z=0.3$, following
the method described in \cite{eisen01a}. 

In both cases, we obtain volume-limited samples by dividing the survey in different galaxy 
populations (according to the absolute magnitude in each case) and then cutting the sample at a minimum and a maximum 
redshift so that the density remains approximately constant. The selected volume-limited samples from the main catalogue are similar
to those used by \cite{zeh05b}, while the LRG one is the same as used by \cite{mart09}.

Finally we restrict the samples to a region of the sky that is nearly complete except for small areas 
masked by bright stars. For this we cut the sample in the survey coordinate system 
$(\eta,\lambda)$ with limits $-31.25 \degre<  \eta < 28.75 \degre$ and $-54.8 \degre< \lambda < 51.8 \degre$.
 Because of this, the samples are smaller and we have less statistics for correlation function estimation, 
but it is simpler for obtaining simulations in the same volume. 

We give in Table~\ref{caract} the magnitude and redshift limits used to construct the four volume-limited samples. We also give
their total number of galaxies ($N_g$), volume ($V$) and mean density ($\bar{n}$).

\begin{table}[H]
{ \footnotesize
\begin{tabular}{|c|c|c|c|c|c|c|}
\hline
Name  & Magnitude Limits & Redshift Limits & Distance Limits      & $N_g$ & $V$                      & $\bar{n}$ \\
      &                  &                 & ($h^{-1}\text{Mpc}$) & 	  & $(h^{-1}\, \text{Mpc})^3$ & $(h^{-1}\text{ Mpc})^{-3}$ \\
\hline
main1 & $M_r < -20$      & $0.038<z<0.119$ & $113.04<d<347.92$    & 127223 & $22.757\times10^6$	      & $5.590\times10^{-3}$  \\
main2 & $M_r < -21$      & $0.059<z<0.168$ & $174.73<d<485.88$    & 67189  & $61.200\times10^6$        & $1.098\times10^{-3}$ \\
main3 & $M_r < -21.5$    & $0.071<z<0.205$ & $209.73<d<587.95$    & 30272 & $108.565\times10^6$       & $2.788\times10^{-4}$ \\
LRG & $-23.544 < M_g < -21.644$ & $0.14<z<0.42$ & $410<d<1140$    & 34347 & $790.4\times10^6$         & $4.345\times10^{-5}$ \\ 
\hline
\end{tabular} }
\caption{\label{caract} Characteristics of the SDSS samples. Distance limits and volume are given for the particular cosmology  $\Omega_M=0.27$, and $\Omega_\Lambda=0.73$.}
\end{table}

%% file: simulations.tex
\subsection{Simulations}
\label{simulations}

\subsubsection{The lognormal model}

The usual paradigm for the distribution of galaxies $n_g$ is the Cox process, i.e. a Poisson process with an intensity given by a continuous field $\rho_g({\bf x})$, which itself is a statistical process. Knowing $\rho_g({\bf x})$ the number of galaxies in a volume $dV$ around ${\bf x}$ is a Poisson variable with intensity $\rho_g({\bf x})dV$. It can be verified that the correlation function of the point process is the same as the underlying continuous process $\rho_g$ plus a weighted Dirac function $ \frac{1}{\bar{n}} \delta_0$ due to the discreteness.

The process $\rho_g$ is linked to the underlying matter density field $\rho_m$ since galaxies form in matter over-densities, but is not supposed to be identical. Indeed it has been observed that correlation is higher in the galaxy distribution than in the matter field, and also depends on galaxy population. The ratio of the two is the square of the mass-luminosity bias $b$. Note that the term bias here has a different meaning than when we speak about the bias of an estimator. The mass-luminosity bias quantifies how fluctuations are amplified in the distribution of galaxies, whereas the bias of an estimator is the difference between its expected value and the quantity to estimate.

In general $b$ should depend on the scale but here we simplify and consider it constant:

\begin{equation}
\xi_g(r)=b^2~ \xi_m(r) .
\label{xibias}
\end{equation}

This simplified model should be a good first order approximation, specially given that we are focusing on the correlation at large scales. This model also takes into account the effect of the peculiar velocities of galaxies in the correlation measurement, known as redshift space distortions. In the simplest plane-parallel approximation, this effect shows as an extra factor multiplying $\xi$ \cite{kaiser87}, which in our case is absorbed in the value of $b$.

We consider a galaxy field $\rho_g$ following a lognormal model as proposed in \cite{cj91}. A lognormal field $Y$ with an expected value of 1 is obtained from a gaussian field $X$ by:

\begin{equation}
\label{relationGLN}
Y=e^{X-\frac{\sigma_X^2}{2}}.
\end{equation}

This model has been successfully applied to density field reconstruction in \cite{kitaura09}, where it enters as a prior model for the matter field. The lognormal model is quite simple and has other interesting properties (see \cite{cj91}): 

\begin{itemize}
\item It describes well the distribution of galaxies as found by Hubble (1934) and recently in \cite{kitaura09} when the galaxy field is smoothed on scales between $10$ and $30$ Mpc
\item The positivity of the field is ensured unlike in a gaussian model
\item Numerous quantities can be calculated as easily as for the gaussian field, e.g. statistics of the peaks, genus
\item It is arbitrarily close to a gaussian field at early times where $\sigma \approx 0$
\item It is the solution of the equations of evolution of $\rho$ when supposing that the initial density field peculiar velocities are gaussian
\end{itemize} 

In the simulations we start by generating the underlying gaussian field and obtain the corresponding lognormal field $Y$ using equation (\ref{relationGLN}). 
The gaussian random field is simple to generate using random Fourier modes ${\bf k}$ that are gaussian with variances $P_G(k)$ (with $P_G$ the underlying gaussian power spectrum). 

For a given power spectrum for the lognormal field $P_{LN}$, we have to know the power spectrum of the underlying gaussian field $P_G$. The relationship between the two fields is simple in terms of covariances (the covariance of the field $Y$ is equal to its correlation function since $\mathbb{E}[Y] = 1$):

\begin{equation}
\label{relationCorrel}
C_G(r) = ln [1 + C_{LN} (r)]	
\end{equation}

The first step is to compute the covariance of the lognormal field $C_{LN}$ from its power spectrum $P_{LN}$ by an inverse Fourier transform in 3 dimensions, i.e. by a Hankel transform in the isotropic case. The power spectrum has bins with exponential sizes in $k$ (i.e. the $ln (k_i)$'s are spaced linearly) since it is smooth in that space. For doing the Hankel transform with this spacing we use the FFTLog progam\footnote{http://casa.colorado.edu/~ajsh/FFTLog/}.
From the lognormal covariance $C_{LN}$, we obtain the gaussian covariance $C_G$ using relationship (\ref{relationCorrel}). Finally the power spectrum of the underlying gaussian field $P_G$ is obtained by a Hankel transform of its covariance $C_G$.

After we have simulated the gaussian field with power spectrum $P_G$ and obtained the lognormal field $Y$ using equation (\ref{relationGLN}), a last step is to adjust the density of the lognormal field, multiplying $Y$ by the expected density $\bar{n}$.

\subsubsection{Adjusting simulation parameters}
\label{parameters}
We adopt, as $P_{LN}$ for our simulations, a $\Lambda \text{CDM}$ power spectrum $P_{\Lambda\text{CDM}}$  given by the iCosmo software \cite{Refregier08} with the following cosmological parameters: $h=0.7, \Omega_b=0.045, \Omega_M=0.27, \Omega_\Lambda=0.73, w_0=-1.0, n_s=1.0, \sigma_8=0.8$. We decided to reproduce the main2 sample, given in section \ref{samples}, which is an average main sample, and the LRG sample. We take the power spectrum at the mean redshift for each sample, i.e. at redshift $z=0.1$ for the main2 sample, and at $z=0.3$ for the LRG sample. 

The simulations give the continuous field $\rho_g$ on a discrete grid of size 700 by 700 by 700 with a physical size of $(1200 ~h^{-1}\,\text{Mpc})^3$ for the main2 sample and $(1600 ~h^{-1}\,\text{Mpc})^3$ for the LRG sample, i.e. with elementary cells respectively of $1.71 ~h^{-1}\,\text{Mpc}$ and $2.29 ~h^{-1}\,\text{Mpc}$. We then place in each cell a number of galaxies which is a Poisson realization of the cell intensity $\rho_g$, with each galaxy placed at random in the cell (i.e. we assume a constant value of $\rho_g$ in each cell).  This will have the effect of smoothing the correlation function approximately with the cell size. The cubic volume is much larger than the final samples but this is done on purpose, since simulations present implicit periodic conditions that create correlations between opposite sides of the cube. We get rid of these correlations when cutting the samples far away from the border.

We choose a mean density of points in the volume that gives on average the same number of points as in the SDSS samples.  

A last step is to choose the mass-luminosity bias $b$ between the samples and the $\Lambda\text{CDM}$ matter correlation function. For estimating this factor we fit the $\Lambda\text{CDM}$ correlation function to the one estimated on the data $\hat{\xi}$:

\begin{equation}
b^2 \, \xi_{\Lambda\text{CDM}} \approx \hat{\xi}
\end{equation}

\begin{figure}[H]
 \begin{center}$
 \begin{array}{cc}
\includegraphics[scale=0.3]{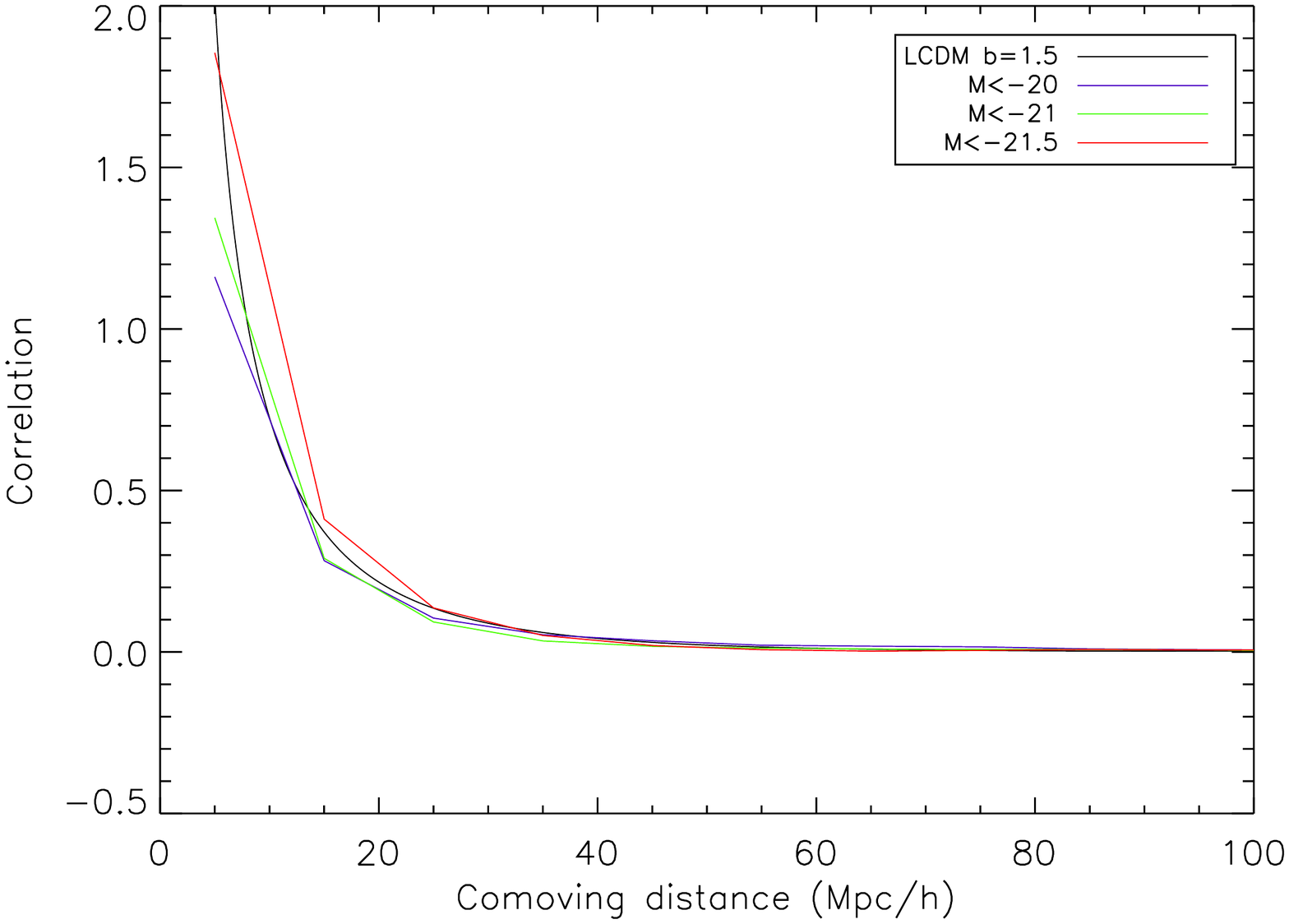}
\includegraphics[scale=0.3]{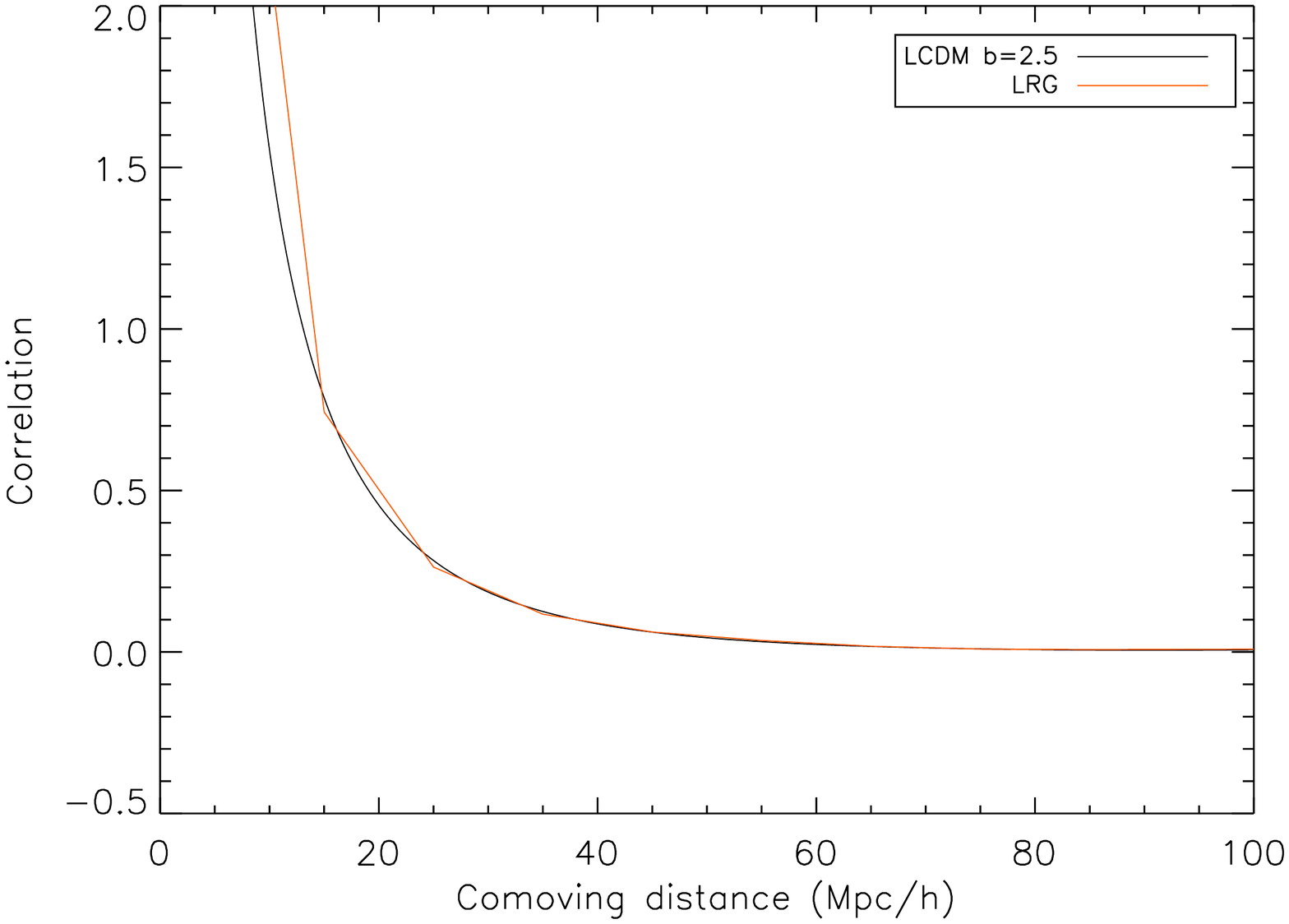}
\end{array}$
\end{center}
\caption{Left Panel: Estimation of the mass-luminosity bias $b$ by fitting the correlation function to $\Lambda\text{CDM}$ correlation. 3 volume-limited samples main1, main2, main3 and $\Lambda\text{CDM}$ correlation at $z=0.1$ with $b=1.5$. Right Panel: LRG sample and $\Lambda\text{CDM}$ correlation at $z=0.3$ with $b=2.5$}
\label{BiasCF}
\end{figure}

By this method we find a bias for the main samples (the variation of $b$ is rather small between the different main samples) and for the LRG sample compared to $\Lambda\text{CDM}$ at redshift $z=0.1$ and $z=0.3$. We find respectively $b=1.5$ and $b=2.5$ (figure \ref{BiasCF}) : as usually observed, the  bias increases with luminosity (\cite{li05},\cite{zeh04}). The bias obtained for the LRG is a bit larger than the one usually found for LRG, $b\approx 2$ (e.g. in \cite{sawan09}). This probably comes from the fact that we selected only brightest galaxies of the LRG population.

%% file: lognormal_2CF.tex
\section{Uncertainty in estimating $\xi$}
\label{lognormal_uncertainty} 

 \subsection{Bias and variance of the estimators}
 \label{comparison}
We use respectively $N=200$ and $N=2000$ lognormal simulations for the main2 and LRG samples with the procedure described before, and compute the different estimators for each realization. We use more simulations for the LRG sample because we want to estimate the covariance matrix of $\xi$ in this sample (see section \ref{compatibility}). 

Each time we use 100 000 random points for computing the estimators (i.e. quantities $RR(r)$ and $DR(r)$ introduced section \ref{2PCF_estimators}). This number is large enough so that the corresponding error is small. Each time a different random catalogue is used, so when we take the mean over all realizations for the analysis of the bias, the effect of finite number of random points is completely negligible. Yet on individual realizations, the fluctuation due to finite number of random points can increase a little bit the variance of the estimators. For a given contribution to the variance, the number of required points is related to the volume size and geometry, and to the size of the bins for estimating $\xi$ (in all our tests we took bins of size $10h^{-1}\text{Mpc}$). More precisely the condition is that $\frac{1}{N_{RR}} RR(r)$ approximates with a given precision $\frac{1} {V^2} \int_V d^3 {\bf x} \int_V d^3 {\bf y} \hspace{2mm} 1_{ |{\bf y-x}| \in [r \pm dr/2] }$. 

\begin{figure}[H]
 \begin{center}$
 \begin{array}{cc}
\includegraphics[scale=0.33]{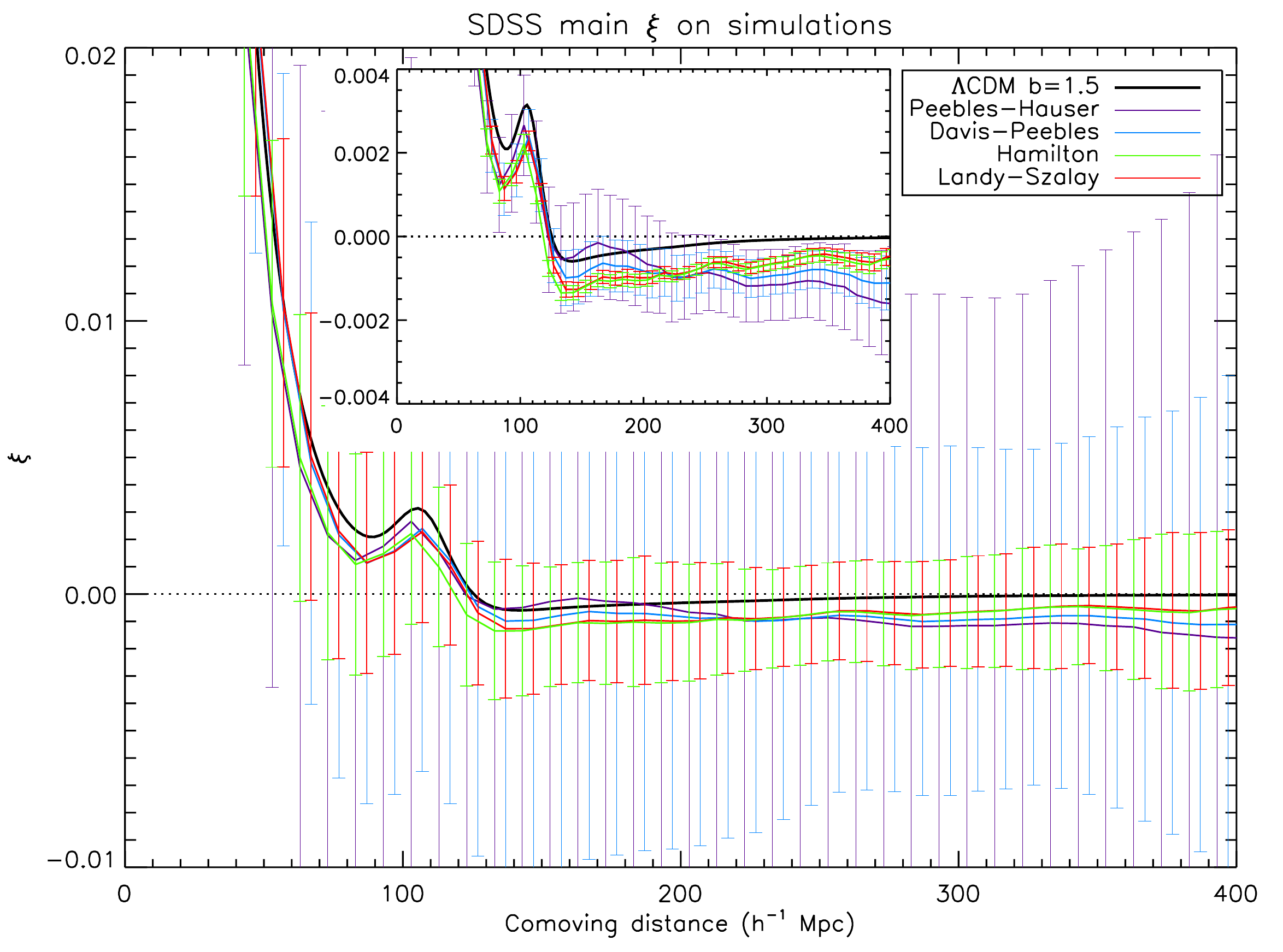} 
\includegraphics[scale=0.33]{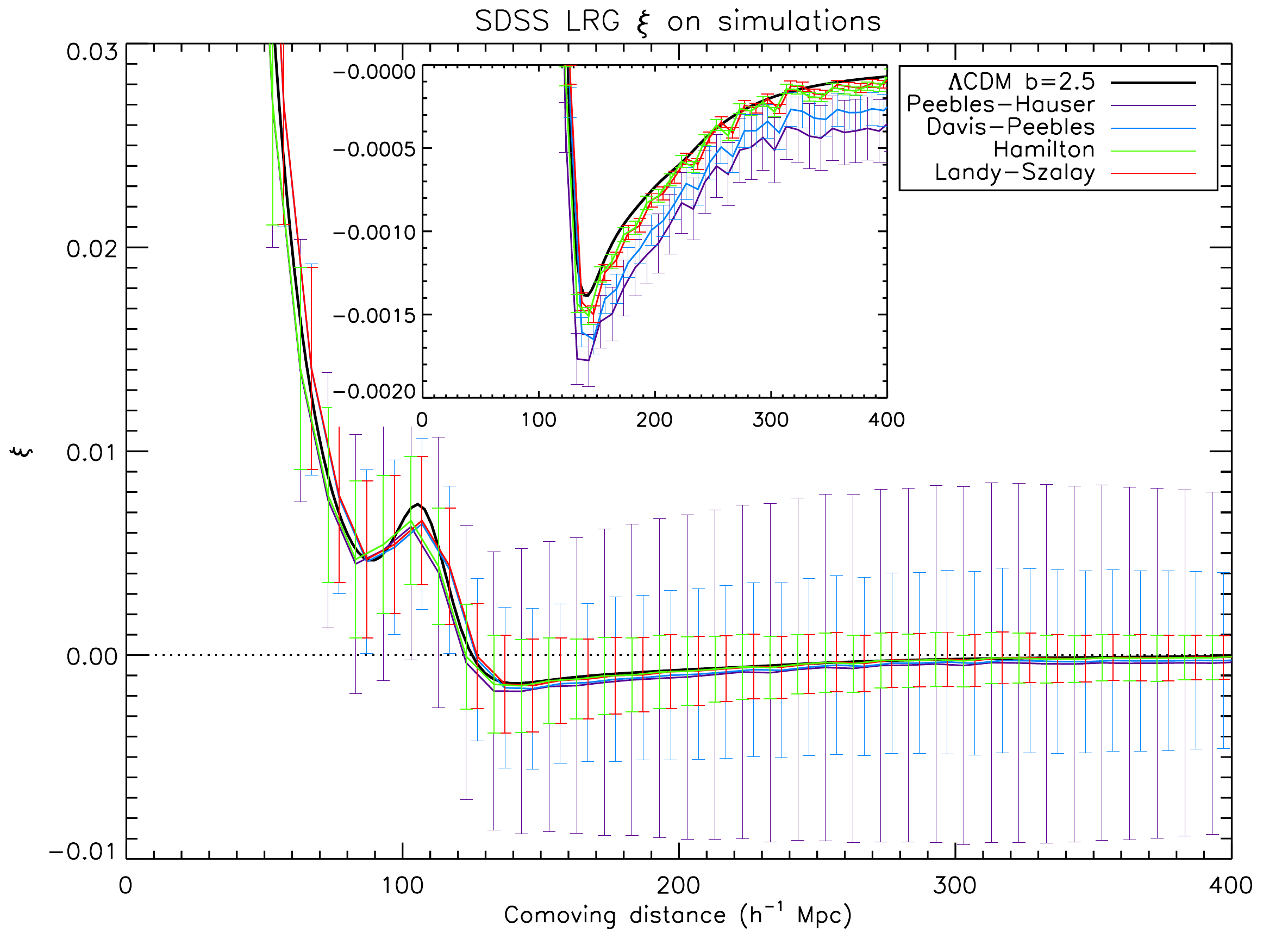}
\end{array}$
\end{center}
\caption{Left Panel: Different estimators' means and $\hat{\sigma}$ on $N=200$ main2 realizations, Peebles-Hauser (purple), Davis-Peebles (light blue), Hamilton (green), Landy-Szalay (red) and $\Lambda\text{CDM}$ at $z=0.1$ with $b=1.5$ (black). Right Panel: Same for the $N=2000$ LRG realizations, except $\Lambda\text{CDM}$ at $z=0.3$ with $b=2.5$. In insets we zoom at the correlation at larger scales. Error bars are divided by $\sqrt{N}$ which gives the uncertainty in the empirical mean on all simulations. This shows the presence of a bias at large scales in the main sample for all estimators. Estimators show a negligible bias in the LRG sample.}
\label{Estimators}
\end{figure}

We show in figure \ref{Estimators} the estimators'  means and standard deviations compared to the theoretical $\Lambda\text{CDM}$ correlation function. For clarity the curves have been translated by $\pm1$ $h^{-1}\text{Mpc}$. A bias can be seen for the estimation in the main sample, with the mean differing by approximately half of the standard deviation from the true value for $r>90 ~h^{-1}\text{Mpc}$. This is shown clearly in the inset where we plot the mean and the uncertainty in the empirical mean on the $N$ simulations, i.e. $\frac{\hat{\sigma}}{\sqrt{N}}$. On the LRG, sample estimators means are nearly indistinguishable from theoretical values.

This also validates our simulation process which gives an output correlation function fitting very well the one in input. There is a small difference at the scale of the BAO (in addition to the bias) that we attribute to the smoothing introduced by grid discretization described section \ref{parameters}. The BAO is a local maximum so the function decreases after smoothing.

Concerning the estimator's variances, they are much smaller on the LRG sample than on the main sample, since the volume is bigger and the Poisson fluctuations remain
small for a number of galaxies $N_D \approx 34000$ and bins of size $10h^{-1}\text{Mpc}$.

We also see that Hamilton and Landy-Szalay estimators are much better than the two others in terms of variance. This agrees with previous studies \cite{pons99}, \cite{ker00} showing a superiority of these estimators on different processes. It also agrees with the analysis in \cite{landy93} considering a Poisson process with no correlation. In the latter case, Landy-Szalay and Hamilton estimators have second order variance decay in $\frac{1}{n^2}$ with $n$ the number of data points  (i.e. a $\frac{1}{|V|^2}$ decay with $|V|$ the volume size) whereas Peebles-Hauser and Davis-Peebles have first order decay in $\frac{1}{n}$. 

\subsection{Effect of the integral constraint}
\label{constraint_data}

We are interested here in the influence of the constraint studied in section \ref{constraint}. The constraint is of the form $\int_0^{r_{max}} f(r) \hat{\xi}(r) \approx 0$ with $f(r) \approx \frac{4 \pi r^2  dr}{|V|}$ for small $r$. Assuming $\int_0^{\infty} r^2 ~\xi(r)~ dr$ is finite, the value of $\int_0^{r_{max}} ~f(r)~ \xi(r)$ vanishes as $\frac{1}{|V|}$ at large volumes.  In usual $\Lambda\text{CDM}$ models the power spectrum verifies $P(0)=0$, and thus the correlation function verifies $\int_0^{\infty} ~r^2 ~ \xi(r) dr=0$, which makes the constraint even more easy to be satisfied. 

Table \ref{constraintSDSS} gives the value of the constrained integral for the theoretical $\xi_{\Lambda\text{CDM}}$  and for the measured $\hat{\xi}$, respectively $I$ and $\hat{I}$. For the main2 sample, $\hat{I}$ is significantly closer to 0 than $I$, meaning that the constraint has an effect on the estimation. The effect is negligible for the LRG sample.

The value of $I$ gives approximately the bias of $\hat{\xi}$ caused by the integral constraint: it is of order $10^{-3}$ for the main2 sample and of order $10^{-4}$ for the LRG sample. Comparing to the values of $\xi$ at the scales of interest (i.e. usually between $50$ and $150 h^{-1} \text{Mpc}$), the bias is significant for the main2 sample but it is negligible for the LRG sample. 

We can also make a parallel with the Cox model of section \ref{constraint}, where the effect of the constraint becomes very small for $\sigma(V) \approx 0.03$. 
The main2 sample has the same value $\sigma(V) \approx 0.03$ but the effect is still important. In the LRG sample the value is 3 times smaller, $\sigma(V)\approx 0.01$, so it is not surprising that the effect is negligible.

\begin{table}[H]
\begin{center}
\begin{tabular}{|c|c|c|c|c|c|c|}
\hline    & $I$ & $\sigma(V)$ & $\hat{I}_{PH}$ & $\hat{I}_{DP}$  & $\hat{I}_{H}$  & $\hat{I}_{LS}$  \\
\hline
main2 &  $6.97 \times 10^{-4}$ & $\approx 0.03$ & $7.52 \times 10^{-6}$ & $-1.20 \times 10^{-4}$ & $-9.70 \times 10^{-5}$ & $4.96 \times 10^{-5}$ \\
\hline
LRG  & $1.68 \times 10^{-4}$ &  $\approx 0.01$ & $-8.04 \times 10^{-5}$ &  $9.14 \times 10^{-6}$ &  $1.12 \times 10^{-4}$ &  $1.25 \times 10^{-4}$ \\  
\hline
\end{tabular}
\caption{Values of $I$ and $\hat{I}$ for different estimators}
\label{constraintSDSS}
\end{center}
\end{table}

\subsection{Reliability of the BAO detection}
\label{reliability}

We use here the Landy-Szalay estimator since we verified it has the lowest variance (with the Hamilton estimator which is nearly equivalent), and like the other estimators has a small negative bias for the main2 sample.

With the $N=200$ main2 simulations and the $N=2000$ LRG simulations we look for the detectability of the BAO in the correlation function, under the form of a bump at about $105~ h^{-1} \,\text{Mpc}$. The situations are different for the main2 (mass-luminosity bias  $b=1.5$) and LRG ($b=2.5$) samples. 
The main2 presents a  lower signal than LRG; and also a larger variance of the estimator due to its smaller  volume.
 
A simple possibility to detect BAOs is to look for a local maximum significantly above 0 in the measured $\hat{\xi}$ for a range of scale around the expected BAO scale, e.g.between 80 and 120 $h^{-1}~\text{Mpc}$. So a simple condition to detect BAOs in most realizations is that the estimator's mean is at more than $1\sigma$ from the 0 level.

\begin{figure}[H]
 \begin{center}$
 \begin{array}{cc}
\includegraphics[scale=0.33]{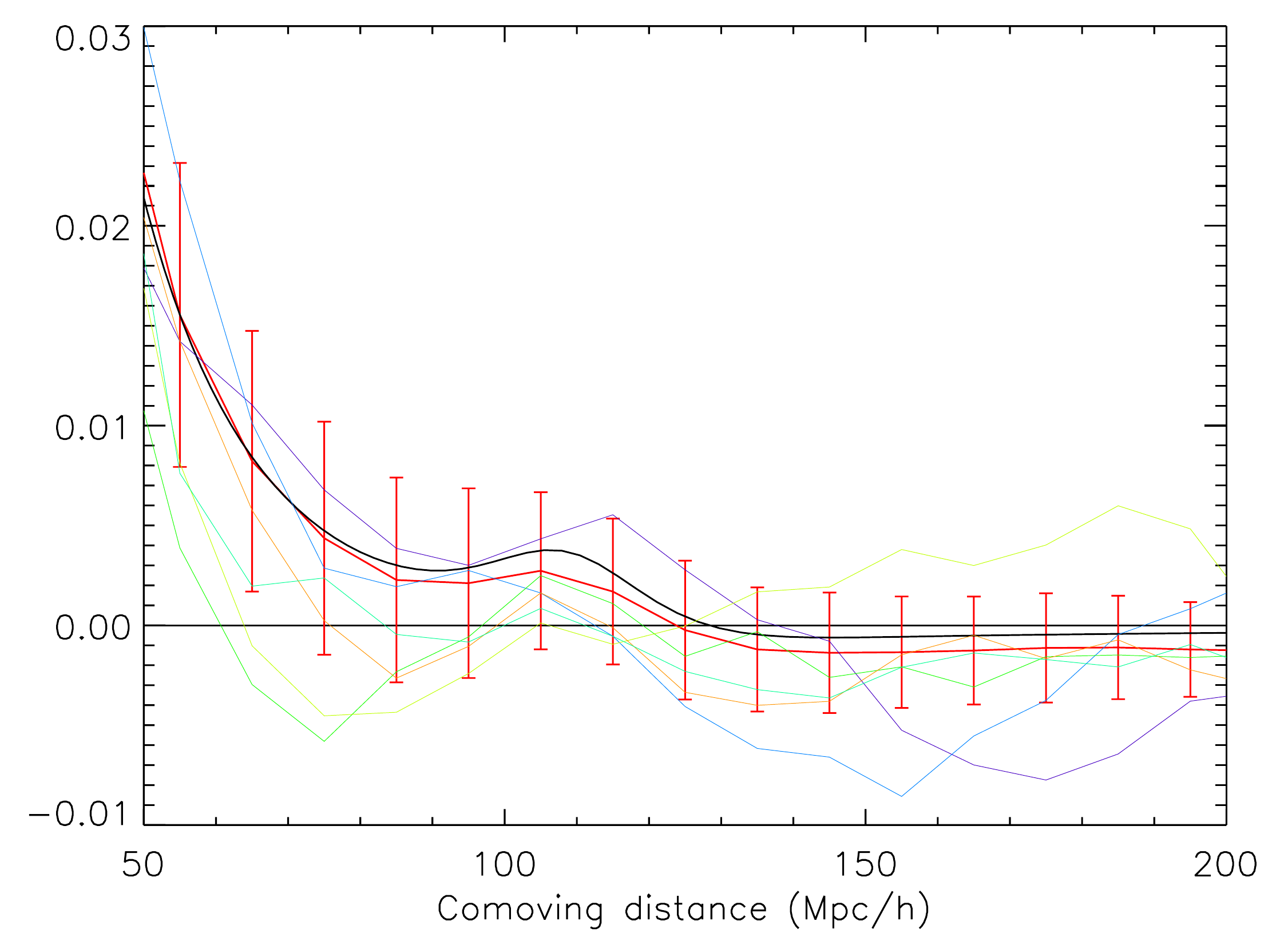}
\includegraphics[scale=0.33]{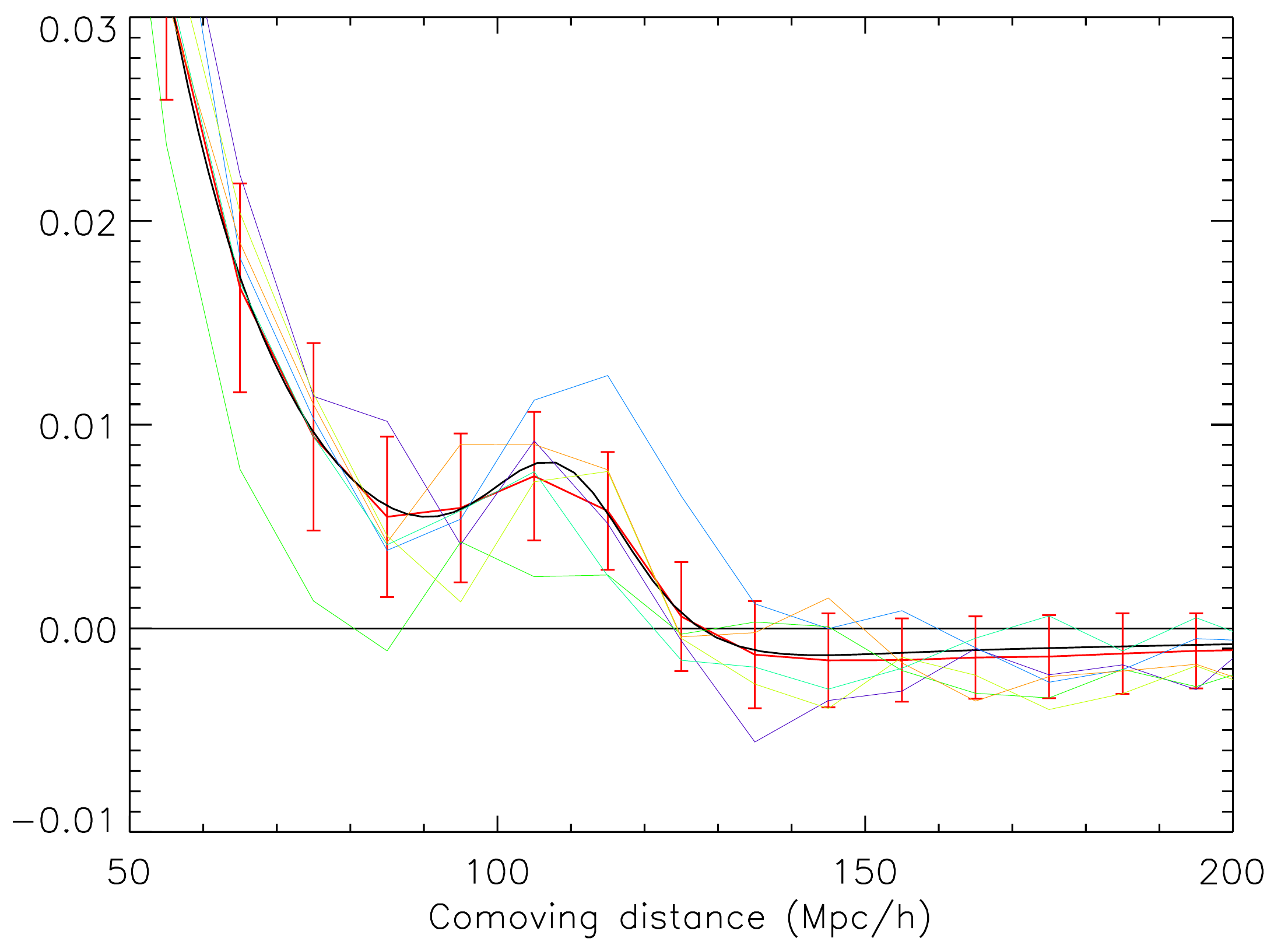}
\end{array}$
\end{center}
\caption{Left Panel: $\hat{\xi}_{LS}$ for 6 lognormal realizations, mean and $1\sigma$ on 200 Main2 realizations (red), and $\xi_{\Lambda\text{CDM}}$  at $z=0.1$ with $b=1.5$ (black). Right Panel: $\hat{\xi}_{LS}$ for 6 lognormal realizations, mean and $1\sigma$ on 2000 LRG realizations (red), and $\xi_{\Lambda\text{CDM}}$  at $z=0.3$ with $b=2.5$ (black)}
\label{6simu}
\end{figure}

This detectability condition is verified for the LRG sample but not for the main2 sample. On figure \ref{6simu}, where we plot $\hat{\xi}_{LS}$ for different realizations, we see a positive BAO peak in the majority of the LRG realizations but less frequently in main2 realizations, where $\hat{\xi}(r)$ is often negative at the peak position.

\subsection{Compatibility to the data}
\label{compatibility}

\begin{figure}[H]
 \begin{center}$
 \begin{array}{cc}
\includegraphics[scale=0.33]{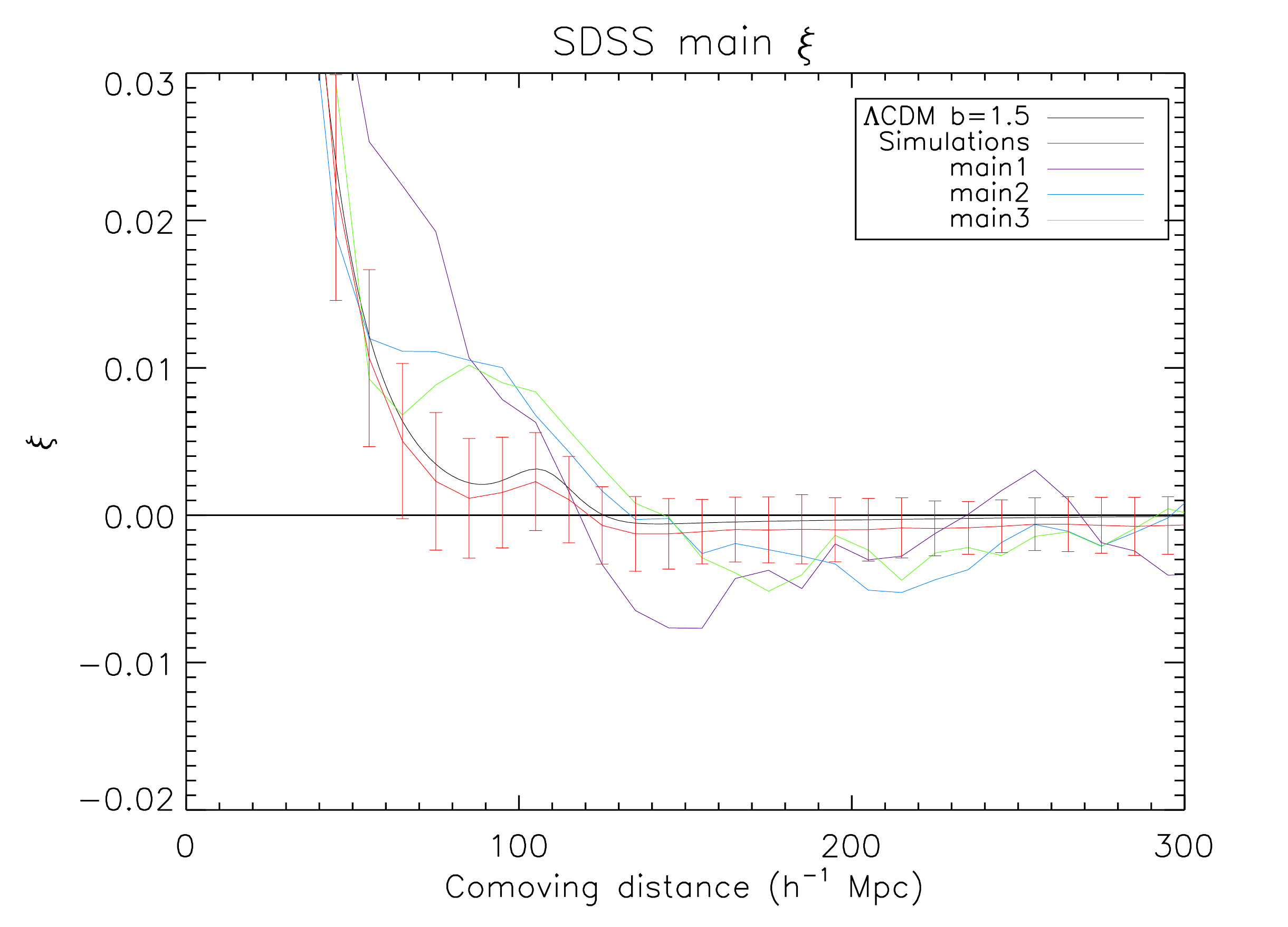}
\includegraphics[scale=0.33]{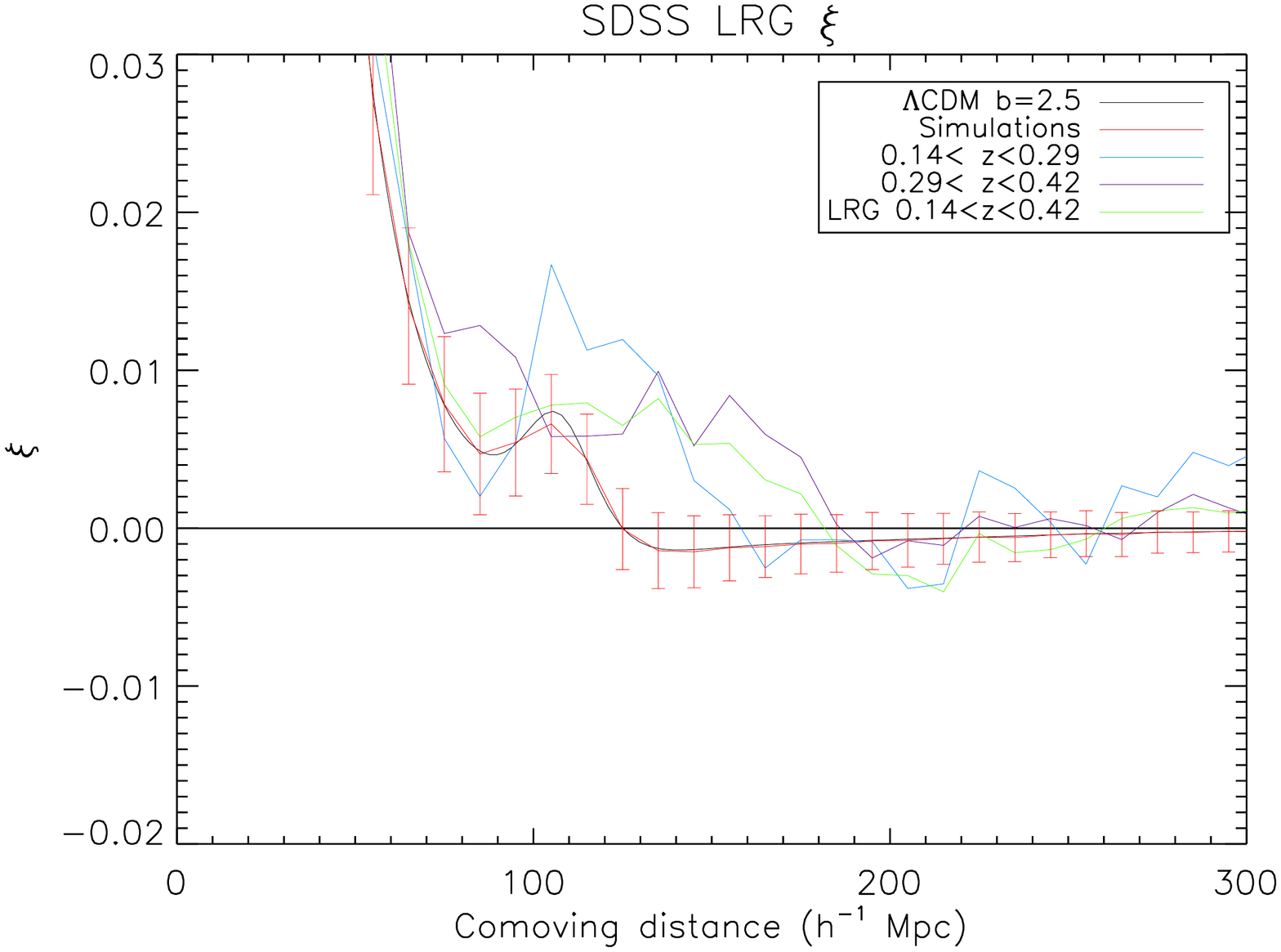}
\end{array}$
\end{center}
\caption{Left panel: data $\hat{\xi}_{LS}$ for different volume-limited samples, main1 (purple), main2 (light blue), main3 (green), mean and $1\sigma$ on 200 lognormal realizations of main2 sample (red) and  $\xi_{\Lambda\text{CDM}}$ at $z=0.1$ with $b=1.5$ (black). Right Panel: data $\hat{\xi}_{LS}$ for LRG volume-limited sample with $0.14<z<0.29$ (light blue), $0.29<z<0.42$ (purple) and the whole sample $0.14<z<0.42$ (green), mean and $1\sigma$ on 2000 lognormal realizations (red) and  $\xi_{\Lambda\text{CDM}}$ at $z=0.3$ with $b=2.5$ (black).}
\label{DataCF}
\end{figure}

We finally compare $\hat{\xi}_{LS}$ of the data samples given in table \ref{caract} with the simulations (figure \ref{DataCF}), keeping in mind that our simulations are not entirely realistic and neglect some systematic effects. For estimating $\xi$ on the SDSS data we took into account the exact survey mask (the angular region observed) when generating random catalogues with the Mangle software\footnote{http://space.mit.edu/$\sim$molly/mangle/}. We explained section \ref{2PCF_estimators} the role of random catalogues in the estimation of $\xi$ that are constructed using the same geometry as the data catalogue. In our study we restricted the data catalogue to the continuous sky region $-31.25 \degre<  \eta < 28.75 \degre$ and $-54.8 \degre< \lambda < 51.8 \degre$, where the SDSS mask is nearly uniform except for small holes caused by bright stars. We found that taking the exact mask into account does not change significantly the results.

For the main samples the estimations are compatible with the simulations. Results of the previous section explain why the BAO peak cannot be seen, except on the main3 sample which is the largest main sample we constructed. 

For the LRG sample, our results agree with previous studies made on the LRG samples of the SDSS DR7, with a less limited angular region and more galaxies (\cite{mart09}, \cite{kazin10}). As in these studies, the BAO peak is much wider than expected:  $\xi$Ê deviates from the $\Lambda\text{CDM}$ value by approximately $3\sigma$ from $140$ to $180 h^{-1} \, \text{Mpc}$ (figure \ref{DataCF}). 

The widening of the peak is more present at higher redshift as can be seen be cutting the LRG sample in 2 redshift ranges (figure \ref{DataCF}). This was already found in \cite{cabre09}, where an analysis for possible systematic effects in the correlation function estimation is done. The conclusion is that none can explain this excess in $\hat{\xi}$. In \cite{kazin10} the sample called DR7-Bright is similar to the one used here and also present an unlikely fit to a particular $\Lambda\text{CDM}$ model.

To quantify the significance of the deviation we follow partly the analysis in \cite{kazin10} and perform a $\chi^2$ test on the correlation function of the whole LRG sample in the range 50 to 200 $h^{-1}~\text{Mpc}$ and in the range 50 to 400 $h^{-1}~\text{Mpc}$. We introduce a new bias parameter free, $\beta$, and we first minimize over $\beta$ the quantity:

\begin{equation}
\chi^2(\beta)=\sum^{N_b}_{i,j=1} [\hat{\xi}(r_i) - \beta \xi_{LN}(r_i)]  C^{-1}_{i,j}  [\hat{\xi}(r_j) - \beta \xi_{LN}(r_j)] 
\label{chi2}
\end{equation}

with $N_b$ the number of bins of the correlation function in the range that we consider.

For a given $\beta$, this is proportional to the log-likelihood assuming a gaussian model for $\hat{\xi}$ with mean $\xi(\beta)=\beta \xi_{LN}$ and covariance matrix $(C_{i,j})$ between bins. In practice $(C_{i,j})$ is estimated with the $N=2000$ lognormal realizations and then inverted. With bins of sizes 10 $h^{-1}~\text{Mpc}$, the covariance matrices are computed respectively on $N_b=15$ bins and $N_b=35$ bins for the analysis in the range 50 to 200 $h^{-1}~\text{Mpc}$ and in the range 50 to 400 $h^{-1}~\text{Mpc}$. This gives respectively 105 and 595 free parameters in the covariance matrix. The number of simulations ($N=2000$) is much greater than this number of parameters in the first case, and also quite larger in the second case, which means the empirical covariance matrix should give a good estimate of the true covariance matrix (see e.g. \cite{pope08}).

With the gaussian hypothesis, $\chi^2(\beta_{min})$ follows a $\chi^2$ law with $N_b-1$ degrees of freedom. We stress that this is only true because of the special way that $\beta$ intervenes in the fitting form $\xi(\beta)$ of equation (\ref{chi2}), and would not be true for any parameter $\theta$ intervening in a fitting form $\xi(\theta)$.

With this procedure we find a value $\chi^2=30.57$ for 14 degrees of freedom in the range 50 to 200 $h^{-1}~\text{Mpc}$, and we find a value $\chi^2=60.86$ for 34 degrees of freedom in the range 50 to 400 $h^{-1}~\text{Mpc}$. These 2 values correspond to $p$-values of respectively $6.3 \times 10^{-3}$ and $3.1 \times 10^{-3}$. Another way to obtain $p$-values without the gaussian hypothesis is to perform the same procedure on the lognormal realizations. For each lognormal realization, we obtain a value $\chi^2(\beta_{min})$, where $\beta_{min}$ is calculated each time to minimize $\chi^2(\beta)$. Among the $N=2000$ realizations we obtain 16 realizations that have higher values of $\chi^2(\beta_{min})$ for the range 50 to 200 $h^{-1}~\text{Mpc}$ and 9 realizations for the range 50 to 400 $h^{-1}~\text{Mpc}$, i.e. we obtain $p$-values of respectively $8 \times10^{-3}$  and $4.5 \times 10^{-4}$. Thus we find an unlikely fit to the particular $\Lambda\text{CDM}$ model used here.

An explanation could be that lognormal simulations do not capture correctly the variance and covariance of the real galaxy distribution. It could also be due to the systematics of the analysis: absence of scale-dependent mass-luminosity bias in the simulations, possibly wrong redshift to distance conversion in the data. Also, with different cosmological parameters in the $\Lambda\text{CDM}$ correlation function, results would have been different, and possibly the deviation less significant.

%% file: conclusion.tex
\section*{Conclusion}
\addcontentsline{toc}{section}{Conclusion}

We have studied uncertainties in correlation function estimators with two different goals: comparing the different estimators on current galaxy surveys (in particular at large scales for BAO study), and study the bias created by the integral constraint.

We simulated  lognormal mock galaxy catalogues; the different parameters of the simulations were adjusted  to those of the SDSS samples: mean redshift of the $\Lambda\text{CDM}$ input power spectrum, density of galaxies in the sample, mass-luminosity bias. Using enough realizations, we quantified the uncertainty in $\xi$ coming from both  estimators' variances and biases.

We first compared the different estimators, in particular regarding their sensitivity to the fluctuation in the number of galaxies $n$ (i.e. the uncertainty in the mean density): Peebles-Hauser  and Davis-Peebles depend at first order on that  fluctuation;  whereas Hamilton and Landy-Szalay have a second order dependence. As a consequence, the variances of the first two estimators have only a first order decay in the volume size, whereas the two latter estimators have a second order decay. We confirmed with  the simulations that Hamilton and Landy-Szalay have much smaller variances.

Then we evaluated the effect of the integral constraint in our simulations: it  can affect the estimation for small volumes, but it becomes negligible when the real $\xi$ itself is close to to verify the constraint. For the Cox process the effect becomes very small when fluctuations in the volume are less than $3\%$ ($\sigma(V)<0.03$). This homogeneity level is achieved for one of the main galaxy sample. Yet for this sample the integral constraint still affects the estimation, with a bias in $\hat{\xi}(r)$ of approximately $0.5 \, \sigma$ for $r>90 ~h^{-1} \text{Mpc}$. For the LRG sample, with  $\sigma(V) \approx 0.01$, estimators are unbiased, thus the integral constraint is not affecting the BAO study.

Finally we were able to determine the reliability of the BAO detection using the estimated correlation function: it is reliable for the LRG sample but not on the main samples. This confirms detections of the BAO signal on the LRG sample we considered and in other studies of $\xi$ on the LRG of SDSS-DR7. However there remains  a large deviation between  $\hat{\xi}$ estimated on the data and our model $\xi_{\Lambda\text{CDM}}$. It consists in a $3\sigma$ deviation from 140 to 180 $h^{-1} \,\text{Mpc}$, which leads to an unlikely fit to our $\Lambda\text{CDM}$ model. The reason for this deviation has not been identified clearly; it could come from systematic effects not taken into account or variance underestimation in the simulations.

\section*{Acknowledgements}
\addcontentsline{toc}{section}{Acknowledgements}

This research was supported by the European Research Council grant ERC-228261 and by the Spanish CONSOLIDER project AYA2006-14056 (including FEDER contributions). PAM acknowledges support from the Spanish Ministerio de Educaci{\'o}n through a FPU contract.

We acknowledge the use of the Sloan Digital Sky Survey data (http://www.sdss.org) and of the NYU Value-Added Galaxy Catalog (http://sdss.physics.nyu.edu/vagc/).

Funding for the SDSS and SDSS-II has been provided by the Alfred P. Sloan Foundation, the Participating Institutions, the National Science Foundation, the U.S. Department of Energy, the National Aeronautics and Space Administration, the Japanese Monbukagakusho, the Max Planck Society, and the Higher Education Funding Council for England. The SDSS Web Site is http://www.sdss.org/.

The SDSS is managed by the Astrophysical Research Consortium for the Participating Institutions. The Participating Institutions are the American Museum of Natural History, Astrophysical Institute Potsdam, University of Basel, University of Cambridge, Case Western Reserve University, University of Chicago, Drexel University, Fermilab, the Institute for Advanced Study, the Japan Participation Group, Johns Hopkins University, the Joint Institute for Nuclear Astrophysics, the Kavli Institute for Particle Astrophysics and Cosmology, the Korean Scientist Group, the Chinese Academy of Sciences (LAMOST), Los Alamos National Laboratory, the Max-Planck-Institute for Astronomy (MPIA), the Max-Planck-Institute for Astrophysics (MPA), New Mexico State University, Ohio State University, University of Pittsburgh, University of Portsmouth, Princeton University, the United States Naval Observatory, and the University of Washington.